# Indentation of a two-dimensional bonded elastic layer with surface tension

Weike Yuan, Gangfeng Wang[*]

*Department of Engineering Mechanics, SVL, Xi'an Jiaotong University, Xi'an 710049, China*

**Abstract:** Surface tension is a prominent factor for the deformation of solids at micro-/nano-scale. This paper investigates the effects of surface tension on the two-dimensional contact problems of an elastic layer bonded to the rigid substrate. Under the plane strain assumption, the elastic field induced by a uniformly distributed pressure within a finite width is formulated by applying the Fourier integral transform, and the limiting process leading to the solutions for a line force brings the requisite surface Green's function. For the indentation of an elastic layer by a rigid cylinder, the corresponding singular integral equation is derived, and subsequently solved by using an effective numerical method based on Gauss-Chebyshev quadrature formula. It is found from the theoretical and numerical results that the existence of surface tension strongly enhances the hardness of the elastic layer and significantly affects the distribution of contact pressure, when the size of contact region is comparable to the elastocapillary length. In addition, an approximated relationship between external load and half-width of contact is generalized in an explicit and concise form, which is useful and convenient for practical applications.

---

[*] Corresponding author. *E-mail address*: wanggf@mail.xjtu.edu.cn

# 1. Introduction

The contact phenomena of soft layers bonded to hard substrate frequently emerge in practical engineering, such as the indentation tests [1-3], the design of textile and paper machinery [4,5] and many other fields in tribology [6]. For many of these situations, the contact can be modelled as a two-dimensional case that an elastic layer of finite thickness is compressed by a rigid cylinder. Thus, the elastic solution of this problem serves the fundament in a wide range of applications.

For the normal contact between elastic solids without friction, the pioneering work presented by Hertz [7], which has been widely used in practice, gives a comprehensive analysis of the elastic field. However, the Hertz theory breaks down when the contact solid is so thin that the influence of substrate must be taken into account. The contact of solids of finite thickness has also been studied in the light of classical elasticity by many researchers. Due to the difficulty in solving the boundary integral equations, instead of exact solutions, some asymptotic solutions were presented for practical applications [5,8,9]. By contrast, the numerical methods can work it out in a simple and direct manner. For instance, the unknown surface tractions between a cylindrical roller and a bonded elastic layer, which were discretized into overlapping triangular elements, can be determined efficiently by solving the matching equations obtained from the boundary conditions of displacements [10,11]. By applying the contour integration to calculate the improper integral of the surface Green's function, Greenwood [12] revisited the contact problem systematically that a rigid cylinder was pressed into an elastic layer which is free to slip on or bonded to the rigid substrate.

As the tremendous development in micro-/nano-electro-mechanical system (MEMS/NEMS), the study of mechanical behaviors of nanobeams or nanowires [13-16], nanoparticles, nanofilms [17,18] and etc. has attracted extensive attention. Owing to the relatively large fraction of surface area to the volume, the surface free energy which is neglected in traditional elasticity plays an important role in the elastic properties of the micro-/nano-sized structures [19] and nanomaterials [20]. The surface elasticity theory developed by Gurtin and Murdoch [21,22] has been regarded as an effective tools in the analysis of mechanical behaviors of nanostructures and nanomaterials, in which surface effects associated with surface tension and surface elasticity are incorporated by modeling the surface as a mathematical layer of zero thickness ideally adhering to the underlying bulk material. Also, the surface elasticity theory has been extended to analyze the micro-/nano-contact problems, when the size of contact is comparable to the elastocapillary length defined as the ratio of surface energy to modulus. By using the double Fourier transform technique, He and Lim [23] derived the three-dimensional surface Green's function for an incompressible elastic half space coupled with surface tension. Based on the two-/three-dimensional surface Green's function with surface tension [24,25], Long et al [26,27] analyzed the contact between a rigid cylinder and an elastic half plane, and a rigid sphere and an elastic half space, respectively. It was found that the existence of surface tension leads to a size-dependent contact response and significantly alters the elastic field comparing with the classical Hertz theory. Moreover, Zhao and Rajapakse [28] studied the effects of surface energy on the elastic field of a two-dimensional and an axisymmetric layer bonded to rigid substrate under a given surface traction, respectively. In their work, the out-of-plane contribution of surface tension to the boundary condition is not considered. Most recently, Rungamornrat et al [29] investigated the axisymmetric contact problems of bonded layer with the effects of both surface tension and surface elasticity. Their numerical results revealed that the contribution of surface tension to the normal displacement and normal stress is dominant over that of surface elasticity if only normal load is applied. Similar conclusion was reported by Vasu and Bhandakkar [30] in their study of a two-dimensional contact problem of a rigid cylinder indenting on an elastic layer-substrate system with surface stresses, and Gao et al [31] also pointed out that the surface tension is the crucial factor that affects the contact stiffness of elastic half plane.

Note that the mechanism of the influence of surface tension is different from that of adhesive interaction between two surfaces which was introduced in the well-known Johnson-Kendall-Roberts model [32] and Derjaguin-Muller-Toporov model [33]. It has been experimentally demonstrated that the effects of surface tension on the adhesive contact are quite significant [34,35]. The corresponding contact analyses of a rigid sphere indenting on an elastic half space incorporating both surface tension and adhesion have been carried out in recent years [36-38].

In consideration of the size of the layered-structures in MEMS/NEMS and the prepared samples in micro-indentation and nano-indentation tests, both the Hertz theory and the conventional contact solutions of the elastic layer in contact with the other solid body are not suitable anymore, and there is an urgent need to develop a contact model including the surface effects, which can be easily understand and conveniently used in practice. In the present paper, we are concerned with the effects of surface tension on the normal frictionless contact between a rigid cylinder and an elastic layer of finite thickness bonded to the rigid substrate, which is still not available in literatures. At first, the elastic field of a bonded layer subjected to a uniform normal load over a finite width is formulated by applying the Fourier integral transform, and as an extreme situation, the Green's function with surface tension is derived for further analysis in the indentation of an elastic layer by a rigid cylinder. A novel numerical method is adopted to solve the integral equation of this contact problem, and an explicit approximated solution as well as some numerical results are presented.

## 2. Surface elasticity theory and the general solutions

2.1 Basic equations

According to the surface elasticity theory, the surface of solid is considered as a negligibly thin membrane with specific material constants different from the bulk material. The basic governing equations in the bulk of solid are the same as those in the classical linear elasticity theory, but the presence of surface stress leads to a set of non-classical equilibrium conditions.

For an isotropic elastic solid in the absence of body forces, the equilibrium and constitutive equations in the bulk are

$$\sigma_{ij,j} = 0, \tag{1}$$

$$\sigma_{ij} = \frac{E}{1+\nu}\left(\varepsilon_{ij} + \frac{\nu}{1-2\nu}\varepsilon_{kk}\delta_{ij}\right), \tag{2}$$

where $\sigma_{ij}$ and $\varepsilon_{ij}$ are the components of stress tensor and strain tensor, $E$ and $\nu$ are the elastic modulus and Poisson's ratio in the bulk material, and $\delta_{ij}$ denotes the Kronecker's delta, respectively. Einstein's summation convention is adopted for repeated Latin indices (1, 2, 3) and Greek indices (1, 2) throughout the paper. The strain tensor is described by the classical linear strain-displacement relationship as

$$\varepsilon_{ij} = \frac{1}{2}\left(u_{i,j} + u_{j,i}\right), \tag{3}$$

where $u_i$ denotes the components of displacement vector in the bulk.

On the isotropic surface, the surface stress tensor $\sigma_{\alpha\beta}^s$ is related to the surface strain $\varepsilon_{\alpha\beta}^s$ by the surface constitutive relations as [21,22]

$$\begin{aligned}\sigma_{\alpha\beta}^s &= \tau_0 \delta_{\alpha\beta} + 2(\mu_s - \tau_0)\varepsilon_{\alpha\beta}^s + (\lambda_s + \tau_0)\varepsilon_{\gamma\gamma}^s \delta_{\alpha\beta} + \tau_0 u_{\alpha,\beta}^s, \\ \sigma_{3\beta}^s &= \tau_0 u_{3,\beta}^s,\end{aligned} \tag{4}$$

with

$$\varepsilon_{\alpha\beta}^s = \frac{1}{2}\left(u_{\alpha,\beta}^s + u_{\beta,\alpha}^s\right), \tag{5}$$

where $\tau_0$ is the residual surface tension, $\mu_s$ and $\lambda_s$ are the two surface elastic constants and $u_\alpha^s$ are the components of surface displacement vector, respectively.

Distinguished from the classical elasticity theory, the surface should satisfy the equilibrium condition in terms of the generalized Young–Laplace equation [39-41], that is

$$\begin{aligned}\sigma_{\alpha\beta,\beta}^s &= \sigma_{ij} n_i n_j^\alpha - t_\alpha, \\ \sigma_{3\beta,\beta}^s &= \tau_0 u_{3,\beta\beta}^s = \sigma_{ij} n_i n_j - t_3,\end{aligned} \tag{6}$$

in which $\mathbf{t}(t_1, t_2, t_3)$ denotes the surface traction vector, and $\mathbf{n}(n_1, n_2, n_3)$ and $\mathbf{n}^\alpha(n_1^\alpha, n_2^\alpha, n_3^\alpha)$ are the unit normal vector to the surface and the tangential vector along the $x_\alpha$ direction.

2.2 General solutions of a bonded elastic layer

Referring to the Cartesian coordinate system ($o$-$xyz$), we consider a two-dimensional elastic layer with thickness $h$ subjected to a surface pressure, as shown in Fig.1, and assume the plane strain conditions with $\varepsilon_{yx}=\varepsilon_{yy}=\varepsilon_{yz}=0$. In this case, the equilibrium equations and Hooke's law in the bulk reduce to

$$\frac{\partial \sigma_{xx}}{\partial x}+\frac{\partial \sigma_{xz}}{\partial z}=0, \quad \frac{\partial \sigma_{zx}}{\partial x}+\frac{\partial \sigma_{zz}}{\partial z}=0, \tag{7}$$

$$\begin{aligned}
\varepsilon_{xx} &= \frac{1-v^2}{E}\left(\sigma_{xx}-\frac{v}{1-v}\sigma_{zz}\right), \\
\varepsilon_{zz} &= \frac{1-v^2}{E}\left(\sigma_{zz}-\frac{v}{1-v}\sigma_{xx}\right), \\
\varepsilon_{xz} &= \frac{1+v}{E}\sigma_{xz},
\end{aligned} \tag{8}$$

with

$$\varepsilon_{xx}=\frac{\partial u}{\partial x}, \quad \varepsilon_{zz}=\frac{\partial w}{\partial z}, \quad \varepsilon_{xz}=\frac{1}{2}\left(\frac{\partial u}{\partial z}+\frac{\partial w}{\partial x}\right), \tag{9}$$

where $u$ and $w$ denote the displacements in the $x$ and $z$ directions, respectively.

In addition, the strain components should satisfy the compatibility condition:

$$\frac{\partial^2 \varepsilon_{xx}}{\partial z^2}+\frac{\partial^2 \varepsilon_{zz}}{\partial x^2}-2\frac{\partial^2 \varepsilon_{xz}}{\partial x \partial z}=0. \tag{10}$$

Define the Airy stress function $\varphi(x,z)$, and the stresses in the bulk can be written as

$$\sigma_{zz}=\frac{\partial^2 \varphi(x,z)}{\partial x^2}, \quad \sigma_{xx}=\frac{\partial^2 \varphi(x,z)}{\partial z^2}, \quad \sigma_{zx}=-\frac{\partial^2 \varphi(x,z)}{\partial x \partial z}, \tag{11}$$

which automatically satisfy the equilibrium equation (i.e. Eq. (7)). Substituting the stresses in terms of Airy stress function into the Hooke's law Eq. (8), the compatibility equation evolves into a bi-harmonic equation for Airy stress function. The Fourier integral transforms are adopted to solve this boundary value problem. The Fourier transformation of the Airy stress function with respect to the coordinate $x$ and its inverse transform can be expressed as

$$\tilde{\varphi}(\xi,z) = \frac{1}{\sqrt{2\pi}} \int_{-\infty}^{\infty} \varphi(x,z) e^{i\xi x} \mathrm{d}x, \quad \varphi(x,z) = \frac{1}{\sqrt{2\pi}} \int_{-\infty}^{\infty} \tilde{\varphi}(\xi,z) e^{-i\xi x} \mathrm{d}\xi, \quad (12)$$

so that the governing partial differential equation reduces to a linear ordinary differential equation which has the general solution:

$$\tilde{\varphi}(\xi,z) = (A+Bz)e^{-|\xi|z} + (C+Dz)e^{|\xi|z}, \quad (13)$$

where $A$, $B$, $C$ and $D$ are functions of $\xi$ and will be determined by applying the boundary conditions.

Therefore, the bulk stress components can be expressed in terms of $\tilde{\varphi}(\xi,z)$ as

$$\sigma_{zz}(x,z) = \frac{1}{\sqrt{2\pi}} \int_{-\infty}^{\infty} -\tilde{\varphi}(\xi,z)\xi^2 e^{-i\xi x} \mathrm{d}\xi,$$

$$\sigma_{xx}(x,z) = \frac{1}{\sqrt{2\pi}} \int_{-\infty}^{\infty} \frac{\partial^2 \tilde{\varphi}(\xi,z)}{\partial z^2} e^{-i\xi x} \mathrm{d}\xi, \quad (14)$$

$$\sigma_{zx}(x,z) = \frac{1}{\sqrt{2\pi}} \int_{-\infty}^{\infty} i\xi \frac{\partial \tilde{\varphi}(\xi,z)}{\partial z} e^{-i\xi x} \mathrm{d}\xi.$$

Using the Hooke's law Eq. (8) and the strain-displacement relationship Eq. (9), the displacements can be derived as

$$u(x,z) = \frac{1-v^2}{E} \frac{1}{\sqrt{2\pi}} \int_{-\infty}^{\infty} \left[ \frac{\partial^2 \tilde{\varphi}(\xi,z)}{\partial z^2} + \frac{v}{1-v} \xi^2 \tilde{\varphi}(\xi,z) \right] \frac{ie^{-i\xi x}}{\xi} \mathrm{d}\xi + C_1,$$

$$w(x,z) = \frac{1-v^2}{E} \frac{1}{\sqrt{2\pi}} \int_{-\infty}^{\infty} \left[ \frac{1}{\xi^2} \frac{\partial^3 \tilde{\varphi}(\xi,z)}{\partial z^3} - \frac{2-v}{1-v} \frac{\partial \tilde{\varphi}(\xi,z)}{\partial z} \right] e^{-i\xi x} \mathrm{d}\xi + C_2, \quad (15)$$

where $C_1$ and $C_2$ are two constants of integration as yet to be determined.

For the considered plane problem, the upper surface of the layer is loaded by a pressure distribution $p(x)$, while the bottom surface is perfectly bonded to the rigid substrate. In this case, the two constants $C_1$ and $C_2$ must be subtracted to ensure that the displacements at the bottom surface are zero all the time. Taking account of the effects of surface stresses, the non-classical boundary conditions can be simplified to

$$-\sigma_{zz}(x,0) - p(x) = \tau_0 \left.\frac{\partial^2 w(x,z)}{\partial x^2}\right|_{z=0},$$
$$-\sigma_{zx}(x,0) = K_s \left.\frac{\partial^2 u(x,z)}{\partial x^2}\right|_{z=0}, \quad (16)$$

on the upper surface ($z=0$), and

$$u(x,h) = w(x,h) = 0, \quad (17)$$

at the interface ($z=h$), where $\tau_0$ is the surface tension independent of surface deformation and $K_s = 2\mu_s + \lambda_s$ is a constant of surface elasticity. In this paper, only normal load is considered and the contact is assumed as frictionless. Thus, the surface tension $\tau_0$ is the dominant factor whilst the surface elasticity $K_s$ can be neglected. For simplicity, only the surface tension is going to be accounted for hereinafter.

Substituting the stresses and displacements [Eqs. (14) and (15)] into the boundary conditions [Eqs. (16) and (17)], one obtains four linear equations to determine the four unknowns $A$, $B$, $C$ and $D$

$$\begin{bmatrix} 2+\dfrac{s|\xi|}{1-v} & \dfrac{s(1-2v)}{1-v} & 2-\dfrac{s|\xi|}{1-v} & \dfrac{s(1-2v)}{1-v} \\ -|\xi| & 1 & |\xi| & 1 \\ |\xi| & h|\xi|-2(1-v) & |\xi|e^{2|\xi|h} & \left[h|\xi|+2(1-v)\right]e^{2|\xi|h} \\ |\xi| & 1-2v+h|\xi| & -|\xi|e^{2|\xi|h} & \left[1-2v-h|\xi|\right]e^{2|\xi|h} \end{bmatrix} \begin{bmatrix} A \\ B \\ C \\ D \end{bmatrix} = \begin{bmatrix} \dfrac{2\tilde{p}(\xi)}{\xi^2} \\ 0 \\ 0 \\ 0 \end{bmatrix}, \quad (18)$$

where

$$s = \frac{2\tau_0}{E^*}, \quad (19)$$

$$\tilde{p}(\xi) = \frac{1}{\sqrt{2\pi}} \int_{-\infty}^{\infty} p(x) e^{ix\xi} dx, \tag{20}$$

$s$ is called the elastocapillary length which indicates the dimension that surface tension can affect, and $E^*=E/(1-v^2)$ is the composite modulus. $\tilde{p}(\xi)$ is the Fourier transformation of the applied pressure $p(x)$. By solving the linear algebraic equations (18), one obtains

$$A = \frac{\tilde{p}(\xi)}{F} A_p, \quad B = \frac{\tilde{p}(\xi)}{F} B_p, \quad C = \frac{\tilde{p}(\xi)}{F} C_p, \quad D = \frac{\tilde{p}(\xi)}{F} D_p, \tag{21}$$

where

$$\begin{aligned}
A_p &= \Lambda_1 e^{2|\xi|h} + 2\xi^2 h^2 - 2|\xi|h + \Lambda_2, \\
B_p &= |\xi|\left(\Lambda_1 e^{2|\xi|h} - 2|\xi|h + 1\right), \\
C_p &= \Lambda e^{-2|\xi|h} + 2\xi^2 h^2 + 2|\xi|h + \Lambda_2, \\
D_p &= -|\xi|\left(\Lambda_1 e^{-2|\xi|h} + 2|\xi|h + 1\right), \\
F &= 2s|\xi|^3 \left[-2|\xi|h + \Lambda_1 \sinh(2|\xi|h)\right] + 2\xi^2 \left[\Lambda_1 \cosh(2|\xi|h) + 2\xi^2 h^2 + \Lambda_2\right],
\end{aligned} \tag{22}$$

with $\Lambda_1=3-4v$, $\Lambda_2=8v^2-12v+5$.

The aforementioned solution of the plane strain problem is general, no matter how the normal pressure distributes and no matter whether the elastic layer is compressible or not. It is worth pointing out that the contact pressure of the indentation of an elastic layer by a rigid cylinder is unknown and the direct evaluation in virtue of the general solutions is impossible. Prior to the indentation case, we will formulate the solutions of the case that a uniform pressure is applied over a finite region in order to derive the surface Green's function.

**3. Uniformly distributed pressure case and the surface Green's function**

In this section, the elastic solutions of a uniform pressure distributing within a finite width and a normal concentrated force acting on the bonded layer with surface tension are formulated in sequence. As Fig. 1 shows, a uniform pressure $p(x)=p_0$ is applied over the region $|x| \leq a$ and its Fourier transformation can be derived as

$$\tilde{p}(\xi) = \frac{1}{\sqrt{2\pi}} \int_{-\infty}^{\infty} p(x) e^{ix\xi} dx = \frac{2 p_0 \sin(a\xi)}{\sqrt{2\pi}\xi}. \tag{23}$$

Hence, the elastic field of the layer incorporating surface tension can be obtained by substituting the determined function $\tilde{\varphi}(\xi,z)$ into the Eqs. (14) and (15). In particular, the shape of the deformed surface is the matter of concern for contact problems [42]. By setting $z=0$, we can formulate the stresses and displacements on the deformed surface. The detailed expressions of the normal displacement and normal bulk stress on the surface are given as follows

$$w(x,0) = \frac{4 p_0}{\pi E^*} \int_0^{\infty} \frac{\sin(a\xi)}{\xi^2} \left[ \Lambda_1 - \frac{2\xi h}{\sinh(2\xi h)} \right] \cos(x\xi) \times \frac{d\xi}{s\xi \left[ \Lambda_1 - \frac{2\xi h}{\sinh(2\xi h)} \right] + \left[ \Lambda_1 \coth(2\xi h) + \frac{2\xi^2 h^2 + \Lambda_2}{\sinh(2\xi h)} \right]}, \tag{24}$$

$$\sigma_{zz}(x,0) = -\frac{2 p_0}{\pi} \int_0^{\infty} \frac{\sin(a\xi)}{\xi} \left[ \Lambda_1 + \frac{2\xi^2 h^2 + \Lambda_2}{\cosh(2\xi h)} \right] \cos(x\xi) \times \frac{d\xi}{s\xi \left[ \Lambda_1 \tanh(2\xi h) - \frac{2\xi h}{\cosh(2\xi h)} \right] + \left[ \Lambda_1 + \frac{2\xi^2 h^2 + \Lambda_2}{\cosh(2\xi h)} \right]}. \tag{25}$$

The integrals in Eqs. (24) and (25) can be numerically calculated (see Appendix A).

Apart from the external load parameters ($p_0$ and $a$) and the bulk material parameters ($E$, $v$), it can be easily seen that the contact response is affected not only by the thickness of layer $h$, but also by the surface tension through the elastocapillary length $s$. When the thickness of the layer tends to be infinity i.e. $h\to\infty$, the obtained stresses and displacements will recover the solutions of the contact of two-dimensional half plane with surface tension [24]. On the other hand, the present derivations can reduce to the conventional solutions of the contact of bonded layer of finite thickness [4] if the surface tension is ignored ($s$=0).

Furthermore, The surface Green's function can be deduced by a limiting process on Eq. (24), in which the resultant force always keeps constant one ($2p_0a$=1) while the contact region diminishes ($a\to 0$), that is

$$G(x) = \frac{2}{\pi E^*} \int_0^\infty \frac{\left[\Lambda_1 - \frac{2\xi h}{\sinh(2\xi h)}\right]\cos(x\xi) \times d\xi}{s\xi^2\left[\Lambda_1 - \frac{2\xi h}{\sinh(2\xi h)}\right] + \xi\left[\Lambda_1 \coth(2\xi h) + \frac{2\xi^2 h^2 + \Lambda_2}{\sinh(2\xi h)}\right]}. \quad (26)$$

Similarly, applying the limiting process on Eq. (25), one can obtain the normal bulk stress distribution on the surface caused by a concentrated unit force as

$$S(x) = -\frac{1}{\pi}\int_0^\infty \frac{\left[\Lambda_1 + \frac{2\xi^2 h^2 + \Lambda_2}{\cosh(2\xi h)}\right]\cos(x\xi)\times d\xi}{s\xi\left[\Lambda_1\tanh(2\xi h) - \frac{2\xi h}{\cosh(2\xi h)}\right] + \left[\Lambda_1 + \frac{2\xi^2 h^2 + \Lambda_2}{\cosh(2\xi h)}\right]}. \quad (27)$$

## 4. Indentation of an elastic layer by a rigid cylinder

Based on the above surface Green's function, we consider the indentation of the elastic layer by a rigid cylinder. As Fig. 2 shows, the layer of thickness $h$ bonded to a rigid substrate is compressed by a cylindrical indenter with radius $R$. When a resultant force $P$ is applied on the cylinder along the $z$-axis, the elastic layer deforms, and it induces an indent depth $d$ and a contact region with half width $a$. Denote the pressure inside the contact region by $p(x')$. Using the solutions under a concentrated unit force Eqs. (26) and (27), one can obtain the normal displacement and normal bulk stress on the surface generated by the contact pressure $p(x')$ as

$$w(x,0) = \int_{-a}^{a} G(|x'-x|) p(x') dx', \tag{28}$$

$$\sigma_{zz}(x,0) = \int_{-a}^{a} S(|x'-x|) p(x') dx', \tag{29}$$

respectively.

In the condition of small deformation, the normal surface displacement within the contact region $|x| \leq a$ can be described as

$$w(x,0) = d - \left(R - \sqrt{R^2 - x^2}\right) \approx d - \frac{x^2}{2R}. \tag{30}$$

Combination of Eq. (28) and (30) leads to a singular integral equation as

$$\int_{-a}^{a} G(|x'-x|) p(x') dx' = d - \frac{x^2}{2R}, \tag{31}$$

which holds within the contact region. Differentiating both sides of Eq. (31) with respect to $x$, one has

$$\int_{-a}^{a} -\frac{\partial G(|x'-x|)}{\partial x} p(x') dx' = \frac{x}{R}. \tag{32}$$

Additionally, the sum of the contact pressure equals to the external load $P$, that is

$$\int_{-a}^{a} p(x') dx' = P. \tag{33}$$

Thus, the indentation of the elastic layer with surface tension by a rigid cylinder is formulated by the singular integral equation (32), together with the constrain condition Eq. (33). Unfortunately, it is almost impossible to derive the analytical solution due to the complexity of the surface Green's function. Instead, an elegant numerical method is adopted to solve this problem. The details are given in Appendix B.

Letting $x=0$ in Eq. (28) or (31), the indent depth, i.e. the normal displacement at the center of contact area, can be written as $d = \int_{-a}^{a} G(|x'|) p(x') \mathrm{d}x'$.

## 5. Results and discussions

By observing the derived equations above, we can find the solutions of the considered contact problems of elastic layer which is bonded to rigid substrate are inextricably related to the Poisson's ratio, the thickness of layer and the surface tension. Note that the presented formulas and methods are applicable for both compressible and incompressible materials. Considering the fact that the Poisson's ratio of most soft materials and biomaterials is close to 0.5 or equal to 0.5, only the results of incompressible layer with surface tension compressed by a uniform pressure and a rigid cylinder are delivered and discussed to illustrate the effects of surface tension and layer thickness in this work.

5.1 Uniformly distributed pressure

For various values of layer thickness, the distribution of normal displacement and normal bulk stress on the upper surface, normalized by $4ap_0/(\pi E^*)$ and $p_0$ respectively, are plotted in Fig. 3. By setting the value of $s/a$ to be unity, it can be clearly found that the normal displacement within the load region will definitely decrease while the absolute value of normal stress will increase if the bonded layer gets thinner. Essentially, this is because the influence of rigid substrate to resist layer deformation will be quite significant for thinner layer. For comparison, the normal bulk stress on the surface of a half plane with *s*/*a*=1.0 [24] is given through the black solid lines in Fig. 3b. When the thickness of layer *h* is much larger than the loading size *a*, the influence of rigid substrate can be neglected and our results will approach to the solutions of half plane.

On the other hand, we illustrate the effects of surface tension on the deformed profile and the normal bulk stress. For several values of the ratio *s*/*a*, Fig. 4 displays the distributions of the normalized normal displacement and normalized normal bulk stress on the surface with *h*/*a*=1.0. From Fig. 4a, we can find the existence of surface tension will stiffen the bonded layer. With the increase of the ratio *s*/*a*, the absolute value of normal displacement decreases on the whole surface. It is also found that the normal bulk stress on the surface is apparently affected by the surface tension. When surface tension is included, the normal bulk stress shows a continuous transition across the contact edge rather than a sudden jump from $-p_0$ to zero predicted by classical elasticity. The larger the ratio *s*/*a* is, the more uniformly the normal bulk stress distributes on the whole surface. Moreover, both the normal displacement and normal bulk stress approaches the classical results as the surface tension decreases.

5.2 Indentation of an elastic layer by a rigid cylinder

In the conventional contact mechanics, the Hertzian contact pressure of the two-dimensional contact of two cylindrical solids is given as [42]

$$p(x) = \frac{2P}{\pi a}\sqrt{1 - \frac{x^2}{a^2}}. \tag{34}$$

Accordingly, the relationship between the load *P* and the half width of contact area is

$$P = \frac{\pi a^2 E^*}{4R}. \tag{35}$$

However, when the thickness of contact solids is comparable to the size of contact region, the contact pressure will derivate from the Hertzian prediction. For the case of a rigid cylinder pressing on an extremely thin ($h/a<0.05$) incompressible bonded layer, the contact pressure follows [42]

$$p(x) = \frac{15P}{16a}\left(1 - \frac{x^2}{a^2}\right)^2, \tag{36}$$

which corresponds with

$$P = \frac{E^* a^5}{30Rh^3}. \tag{37}$$

For the general situation, it has been reported by Greenwood [12] that the contact pressure will develop from the Hertzian distribution Eq. (34) to Johnson's limiting distribution Eq. (36) as the ratio $h/a$ decreases from a very large value (e.g. $h/a=2.0$) to a much smaller one. Each of the two limiting pressure distribution has a continuous variation and falls to zero at the contact edge $x = \pm a$.

When the size of contact region comes into micro-/nano-scale, the presence of surface tension will also significantly alter the contact pressure. According to the results of the indentation of elastic half plane with surface tension by a rigid cylinder [26], the contact pressure still varies continuously, but tends to be a uniformly distribution and has a positive nonzero value at the contact edge. In this work, we demonstrate the effects of both layer thickness and surface tension on the contact pressure, and further on the surface deformation as well as the normal bulk stress on the surface. Consider a soft incompressible silicone layer ($E$=250 kPa, $\tau_0$=39 mN/m$^2$ [35,36] and thus the elastocapillary length $s$=234 nm according to Eq. (19)) indented by a cylinder with radius $R$=10 μm. Applying the numerical method in Appendix B, the contact pressure can be obtained, and then the normal displacement and normal bulk stress on the surface can be calculated for various ratios $s/a$ and $h/a$. It is noted that the range of $h/a$ is [0.3, 5.0] and the range of $s/a$ is [0.1, 5.0] in the present numerical calculations.

At first, setting the ratio $s/a$=1.0, Fig. 5 shows the distribution of contact pressure for different values of $h/a$. For a thicker elastic layer (e.g. $h/a$>2.0), the obtained contact pressure approaches to the solution of Long et al [26], which means that the influence of layer thickness can be neglected and the assumption of half plane would be reasonable. However, as the thickness of elastic layer is comparable with or smaller than the half-width of contact, the pressure distribution will deviate from that of half plane. Such development of pressure distribution with the variance of the ratio $h/a$ is in qualitatively accord with the results of Greenwood [12], though the surface tension is included. The corresponding normal displacement and normal bulk stress on the surface are plotted in the Fig. 6, which experiences a similar development tendency as that of the uniform pressure case in Fig. 3. On the other hand, Fig. 7 displays the distribution of contact pressure for different values of $s/a$ when $h/a$ keeps a constant value one. We can find that the contact pressure approaches to a uniformly distributed pressure as the ratio $s/a$ increases, and, in contrast, reduces to the classical result as the surface tension is vanished. Correspondingly, Fig. 8 shows the distribution of the normal displacement and normal bulk stress, which qualitatively demonstrates similar characteristics as that of the uniform pressure case in Fig. 4.

Finally, it is instructional to give the relationship between load and contact width. According to the asymptotic solution of Meijers, the load normalized by the Hertzian solution only depends on the ratio of layer thickness to the half-width of contact region, i.e. $P/P_H = f(h/a)$ with $P_H = \pi a^2 E^* / (4R)$ termed as Hertzian load. Calculating the total load for a number of values of $h/a$ when surface tension is neglected, we find that a relatively simple relation can be given by fitting the numerical results, as shown in Fig. 9. In the form of power function, the relationship is

$$\frac{P}{P_H} = 1 + 0.900 \left(\frac{h}{a}\right)^{-1.77}, \tag{38}$$

which approaches to Johnson's approximate expression Eq. (37) as $h/a$ decreases, and reduces to the Hertzian equation Eq. (35) if $h/a$ is larger than 5.

However, when surface tension is included, the normalized load $P/P_H$ should depend not only on $h/a$ which indicates the influence of layer thickness, but also on the ratio $s/a$ that denotes the influence of surface tension. We further modify the normalized load as $P/P_{HL}$ where $P_{HL}$ termed as Meijers' load is expressed as $P_H \left[1 + 0.900 (h/a)^{-1.77}\right]$. Fig. 10 shows the variations of the modified normalized load $P/P_{HL}$ with respect to $s/a$. For an arbitrary given value of $h/a$, the relationship between $P/P_{HL}$ and $s/a$ can be fitted well in the form of power function by

$$\frac{P}{P_{HL}} = 1 + k_1 \left(\frac{s}{a}\right)^{k_2}, \tag{39}$$

where $k_1$ and $k_2$ are parameters to be determined by fitting lots of numerical results. It can be seen that the two parameters are related to the value of $h/a$ and become constants as $h/a$ is larger than 2.0. For $h/a \in [0.3, 2.0]$, the variations of $k_1$ and $k_2$ with respect

to *h*/*a* are plotted in Fig. (11) and can be expressed approximately as

$$k_1 = \frac{0.598 + 2.77\frac{h}{a}}{1 + 0.559\frac{h}{a} + 0.244\left(\frac{h}{a}\right)^2}, \quad k_2 = \frac{0.269 + 4.76\frac{h}{a}}{1 + 5.46\frac{h}{a} - 0.108\left(\frac{h}{a}\right)^2}, \qquad (40)$$

and when $h/a > 2.0$, $k_1 = 2.00$ and $k_1 = 0.850$. In this way, we can present the relationship between load and contact width explicitly for an elastic layer with surface tension, which can be used for micro-/nano-indentation tests conveniently.

## 6. Conclusions

In this paper, we analyze the two-dimensional contact problems of a uniformly distributed pressure, a concentrated line force and a rigid cylinder acting on an elastic bonded layer with surface tension under the plane strain condition, respectively. The normal displacement and normal bulk stress on the surface, as well as the contact pressure for indentation by a rigid cylinder, are presented in details. In general, both surface tension and layer thickness affect the contact response. For a given elastic layer with specified material constants and surface tension, the impact of layer thickness on the layer deformation and contact stress is similar to the classical results. When the size of contact region is on the same magnitude with the elastocapillary length, the effects of surface tension must be taken into account since it significantly alters the contact pressure and the elastic field. Compared with the classical contact solution, the contact pressure of an elastic layer with surface tension tends to be a uniformly distributed as the half-width contact area shrinks. Surface tension decreases the contact deformation, and flattens the distribution of normal bulk stress on the surface, which varies continuously across the contact edge and has nonzero values outside the contact area. Moreover, through a series of numerical calculations for different values of $h/a$ and $s/a$, we find an empirical expression of the half-width of contact related to the indentation load, incorporating the effects of surface tension together with the layer thickness. All the results reveal that the indentation of elastic layer is size-dependent, especially at micro-/nano-scale, and the existence of surface tension makes the layer perform stiffer than that at macro-scale.

**Acknowledgment**

Support from the National Natural Science Foundation of China (Grant No. 11525209) is acknowledged.

**Appendix A:** Numerical evaluation of the normal bulk stress and displacement

The normal bulk stress and normal displacement on the surface caused by the uniform pressure $p_0$ are evaluated by numerical integral method. For convenience, the two equations (24) and (25) are further normalized as

$$\frac{\sigma_{zz}(r,0)}{p_0} = -\frac{2}{\pi}\int_0^\infty \frac{\sin(t)}{t}\frac{\left[\Lambda_1 + \frac{2t^2h'^2 + \Lambda_2}{\cosh(2th')}\right]\cos(rt) \times dt}{s't\left[\Lambda_1\tanh(2th') - \frac{2th'}{\cosh(2th')}\right] + \left[\Lambda_1 + \frac{2t^2h'^2 + \Lambda_2}{\cosh(2th')}\right]}, \quad (A.1)$$

$$\frac{\pi E^* w(r,0)}{4ap_0} = \int_0^\infty \frac{\sin(t)}{t^2}\frac{\left[\Lambda_1 - \frac{2th'}{\sinh(2th')}\right]\cos(rt) \times dt}{s't\left[\Lambda_1 - \frac{2th'}{\sinh(2th')}\right] + \left[\Lambda_1\coth(2th') + \frac{2t^2h'^2 + \Lambda_2}{\sinh(2th')}\right]}, \quad (A.2)$$

where $\xi = \frac{t}{a}$, $h = ah'$, $s = as'$, $x = ar$.

Afterwards, the improper integrals are calculated through the inbuilt numerical integral function *integral* in MATLAB. It should be pointed out that the upper limit of integral in Eqs. (A.1) and (A.2) is set as 10000 rather than infinity, and the accuracy has been checked. This numerical evaluation process is adopted again for the calculation of improper integrals in Appendix B.

**Appendix B:** Numerical method for the singular integral equations

The considered contact problem in this work reduces to solve the singular integral equation Eq. (32) and its constraint condition Eq. (33). In complete forms, they are

$$\frac{1}{\pi}\int_{-a}^{a}\int_{0}^{\infty}\frac{\left[\Lambda_{1}-\dfrac{2\xi h}{\sinh(2\xi h)}\right]\sin\left[(x-x')\xi\right]p(x')}{s\xi\left[\Lambda_{1}-\dfrac{2\xi h}{\sinh(2\xi h)}\right]+\left[\Lambda_{1}\coth(2\xi h)+\dfrac{2\xi^{2}h^{2}+\Lambda_{2}}{\sinh(2\xi h)}\right]}\mathrm{d}\xi\mathrm{d}x'=\frac{E^{*}x}{2R}, \quad (\text{B.1})$$

$$\int_{-a}^{a}p(x')\mathrm{d}x'=P. \quad (\text{B.2})$$

If the thickness of layer tends to be infinity i.e. $h\to\infty$, Eq. (B.1) reduces to that of the indentation of a half plane with surface tension as [26]

$$\frac{1}{\pi}\int_{-a}^{a}\int_{0}^{\infty}\frac{\sin\left[(x-x')\xi\right]p(x')}{s\xi+1}\mathrm{d}\xi\mathrm{d}x'=\frac{E^{*}x}{2R}, \quad (\text{B.3})$$

in which the improper integral with respect to $\xi$ can be analytically deduced as

$$\begin{aligned}I_{half\text{-}plane}(x,x')&=\int_{0}^{\infty}\frac{\sin\left[(x-x')\xi\right]}{s\xi+1}\mathrm{d}\xi\\&=\frac{1}{s}\left[\frac{\pi}{2}\operatorname{sgn}\left(\frac{x-x'}{s}\right)\cos\left(\frac{x-x'}{s}\right)+\sin\left(\frac{x-x'}{s}\right)\operatorname{Ci}\left(\frac{|x-x'|}{s}\right)\right.\\&\quad\left.-\cos\left(\frac{x-x'}{s}\right)\operatorname{Si}\left(\frac{x-x'}{s}\right)\right],\end{aligned} \quad (\text{B.4})$$

where Ci(x) and Si(x) are the cosine and sine integral functions defined as

$$\operatorname{Ci}(x)=-\int_{x}^{\infty}\frac{\cos\phi}{\phi}\mathrm{d}\phi \text{ for } x>0, \quad \operatorname{Si}(x)=\int_{0}^{x}\frac{\sin\phi}{\phi}\mathrm{d}\phi \text{ for } -\infty<x<\infty, \quad (\text{B.5})$$

and

$$\operatorname{sgn}(x)=\begin{cases}1, & x>0\\0, & x=0,\\-1, & x<0\end{cases} \quad (\text{B.6})$$

is the sign function. When surface tension is neglected (i.e. $s=0$), the integral equations Eq. (B.1) and Eq. (B.3) reduce to that of the conventional contact problem of an elastic layer, and that of classical Hertzian contact (see ref. [5] and [42]), respectively.

However, the improper integral in Eq. (B.1) is too complex to deduce analytically. Alternatively, the numerical evaluation process in Appendix A is employed and the

convergence issue can be avoided by rewriting this improper integral [30] as

$$I_{layer}(x,x') = \int_0^\infty \frac{\left[\Lambda_1 - \frac{2\xi h}{\sinh(2\xi h)}\right]\sin[(x-x')\xi]}{s\xi\left[\Lambda_1 - \frac{2\xi h}{\sinh(2\xi h)}\right] + \left[\Lambda_1 \coth(2\xi h) + \frac{2\xi^2 h^2 + \Lambda_2}{\sinh(2\xi h)}\right]} d\xi$$

$$= \int_0^\infty \left\{ \frac{\left[\Lambda_1 - \frac{2\xi h}{\sinh(2\xi h)}\right]\sin[(x-x')\xi]}{s\xi\left[\Lambda_1 - \frac{2\xi h}{\sinh(2\xi h)}\right] + \left[\Lambda_1 \coth(2\xi h) + \frac{2\xi^2 h^2 + \Lambda_2}{\sinh(2\xi h)}\right]} - \right. \quad \text{(B.7)}$$

$$\left. \frac{\sin[(x-x')\xi]}{s\xi + 1} \right\} d\xi + I_{half\text{-}plane}(x,x').$$

Setting $\xi = \frac{t}{a}$, $h = ah'$, $s = as'$, $x = ar$, $x' = ar'$, $p(r') = \frac{P}{2a}p'(r')$, Eqs. (B.1) and

(B.2) can be normalized as

$$\frac{1}{\pi}\int_{-1}^1 I'_{layer}(r,r')q(r')\frac{dr'}{\sqrt{1-r'^2}} = \frac{E^*a^2 r}{PR}, \quad \text{(B.8)}$$

$$\frac{1}{\pi}\int_{-1}^1 q(r')\frac{dr'}{\sqrt{1-r'^2}} = \frac{2}{\pi}, \quad \text{(B.9)}$$

where $q(r') = p'(r')\sqrt{1-r'^2}$ and

$$I'_{layer}(r,r') = \int_0^\infty \left\{ \frac{\left[\Lambda_1 - \frac{2th'}{\sinh(2th')}\right]\sin[(r-r')t]}{s't\left[\Lambda_1 - \frac{2th'}{\sinh(2th')}\right] + \left[\Lambda_1 \coth(2th') + \frac{2t^2 h'^2 + \Lambda_2}{\sinh(2th')}\right]} - \right. \quad \text{(B.10)}$$

$$\left. \frac{\sin[(r-r')t]}{s't + 1} \right\} dt + I'_{half\text{-}space}(r,r'),$$

with

$$I'_{half\text{-}space}(r,r') = \frac{1}{s'}\left[\frac{\pi}{2}\text{sgn}\left(\frac{r-r'}{s'}\right)\cos\left(\frac{r-r'}{s'}\right) + \sin\left(\frac{r-r'}{s'}\right)\text{Ci}\left(\frac{|r-r'|}{s'}\right) \right.$$

$$\left. -\cos\left(\frac{r-r'}{s'}\right)\text{Si}\left(\frac{r-r'}{s'}\right)\right]. \quad \text{(B.11)}$$

Using the Gauss-Chebyshev quadrature formula, the two integral equations (B.8)

and (B.9) are transformed into the following linear algebraic equations [43]

$$\frac{1}{n}\sum_{j=1}^{n} I'_{layer}\left(r_i, r'_j\right) q\left(r'_j\right) = \frac{E^* a^2 r_i}{PR},$$
$$\frac{1}{n}\sum_{j=1}^{n} q\left(r'_j\right) = \frac{2}{\pi},$$
(B.12)

where

$$r_i = \cos\left(\frac{i}{n}\pi\right), \quad i = 1, 2, \cdots, n-1,$$
$$r'_j = \cos\left(\frac{2j-1}{2n}\pi\right), \quad j = 1, 2, \cdots, n.$$
(B.13)

For a specific indentation case, the external load and the half-width of contact area should be given simultaneously prior to the determination of contact pressure by solving the linear equations (B.12). Hence, a guess-iteration method is applied, which can be achieved by the following path: preset half-width of contact and iteratively guess load to solve linear equations for contact pressure. If the trial value of load is much smaller than the actual one, the pressure at the fringe would be far less than that inside; conversely, if the trial value is much larger than the actual one, the pressure would be extremely large at the fringe. The final obtained pressure should be continuous within the whole contact region, which corresponds to the actual real value of external load.

**Figure captions:**

Fig. 1. Contact schematic of an elastic layer under a uniform pressure over a finite width.

Fig. 2. Contact between a two-dimensional elastic layer and a rigid circular cylinder.

Fig. 3. Surface elastic field of the bonded elastic layer under a uniformly distributed pressure for $s/a=1$: (a) distribution of the normal displacement normalized by $4ap_0/(\pi E^*)$ and (b) distribution of the normal bulk stress normalized by $p_0$.

Fig. 4. Surface elastic field of the bonded elastic layer under a uniformly distributed pressure for $h/a=1$: (a) distribution of the normal displacement normalized by $4ap_0/(\pi E^*)$ and (b) distribution of the normal bulk stress normalized by $p_0$.

Fig. 5. The contact pressure distribution for $s/a=1$.

Fig. 6. Surface elastic field of a bonded elastic layer indented by a rigid cylinder for $s/a=1$: (a) distribution of the normal displacement normalized by $P/E^*$ and (b) distribution of the normal bulk stress normalized by $P/(2a)$.

Fig. 7. The contact pressure distribution for $h/a=1$.

Fig. 8. Surface elastic field of a bonded elastic layer indented by a rigid cylinder for $h/a=1$: (a) distribution of the normal displacement normalized by $P/E^*$ and (b) distribution of the normal bulk stress normalized by $P/(2a)$.

Fig. 9. Variation of the normalized load $P/P_H$ with respect to the ratio $h/a$ without surface tension.

Fig. 10. Variation of the normalized load $P/P_{HL}$ with respect to the ratio $s/a$ for various values of $h/a$. (Scatters: numerical results; Lines: fitting curves)

Fig. 11. Dependence of the fitting parameters on the ratio $h/a$: (a) $k_1$ and (b) $k_2$.

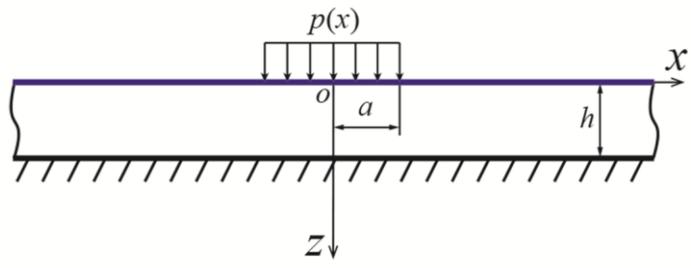

Figure 1

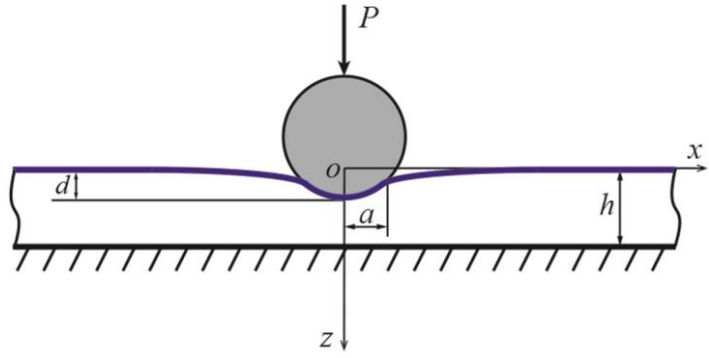

Figure 2

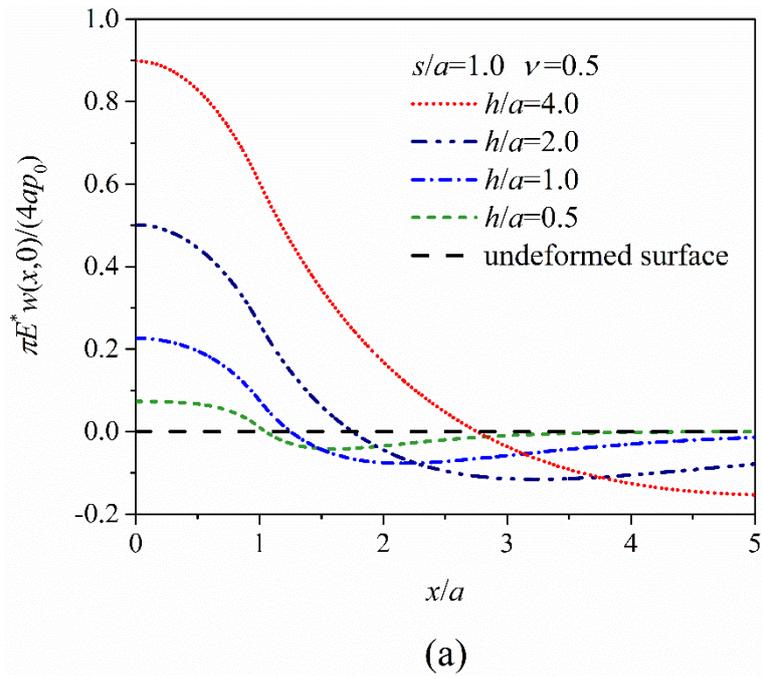

(a)

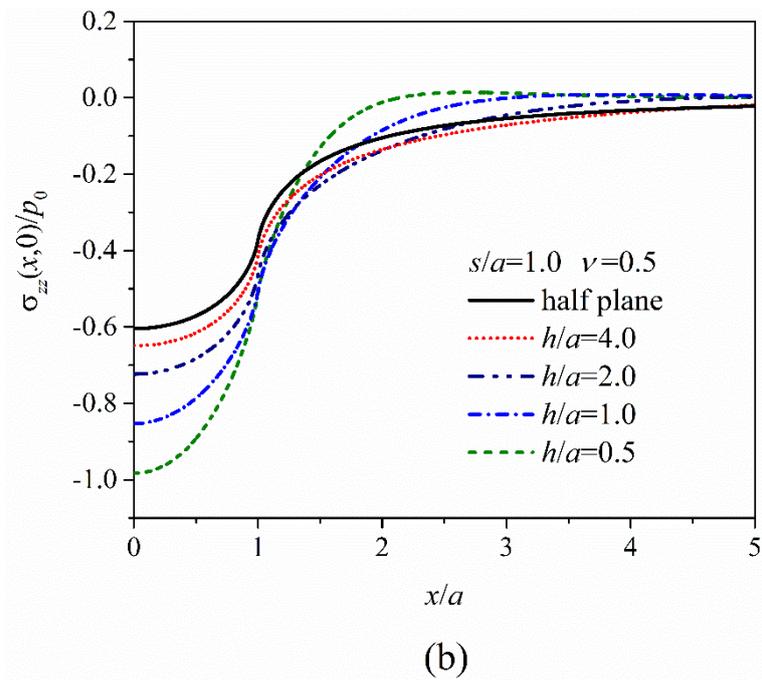

(b)

Figure 3

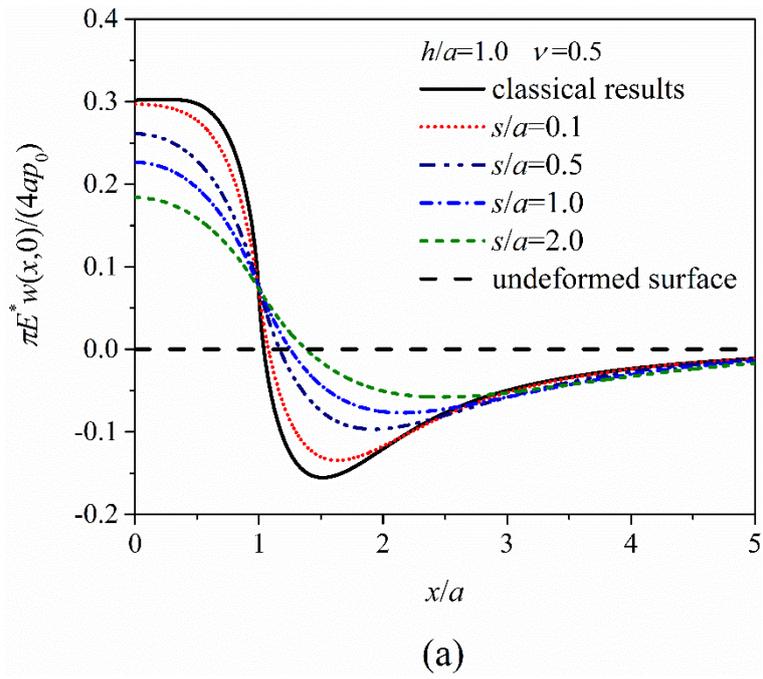

(a)

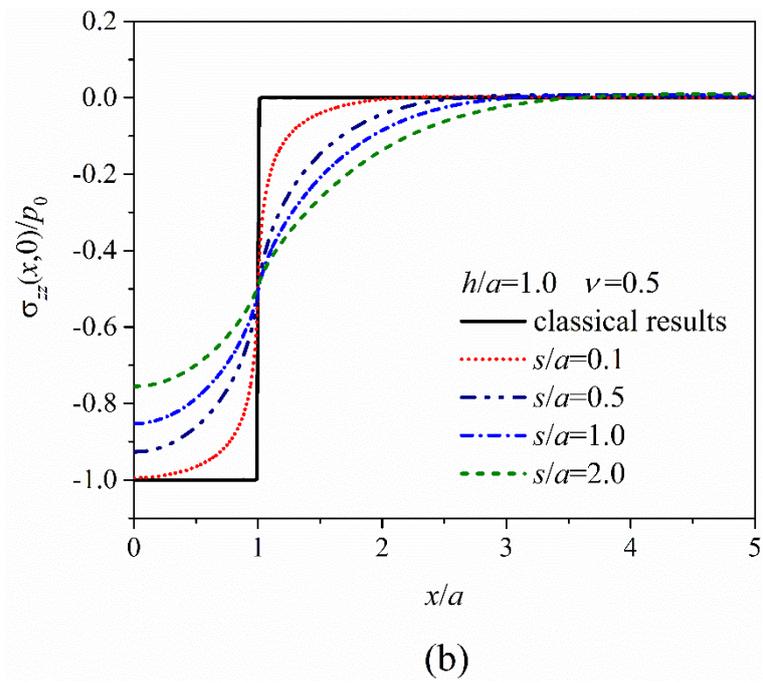

(b)

Figure 4

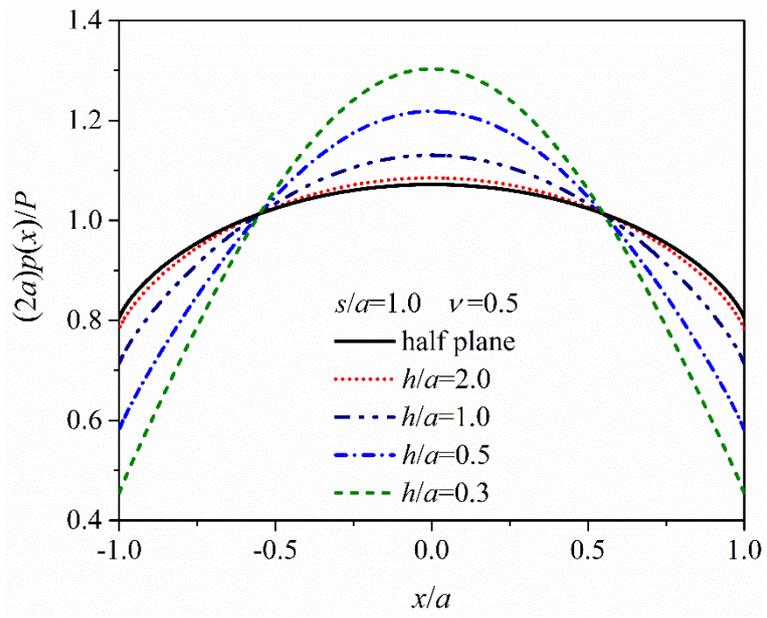

Figure 5

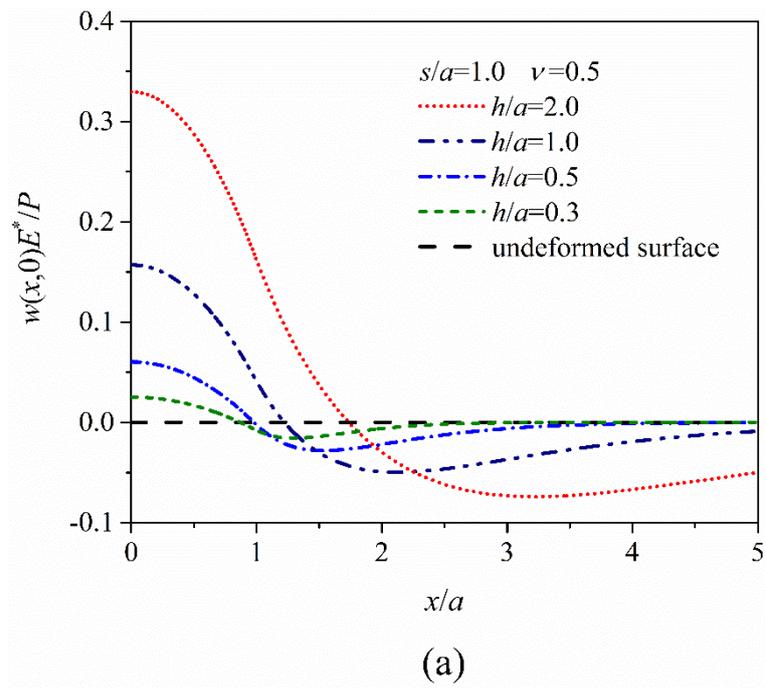

(a)

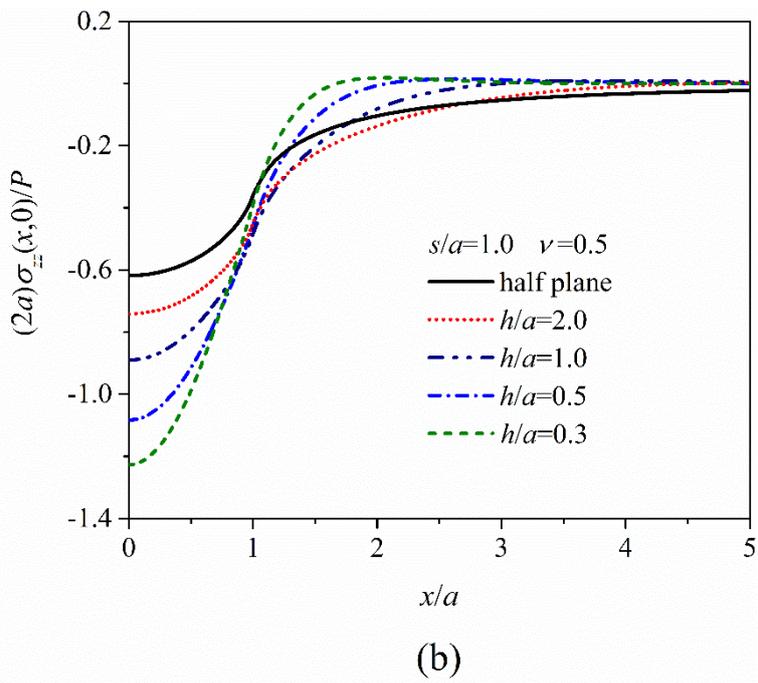

(b)

Figure 6

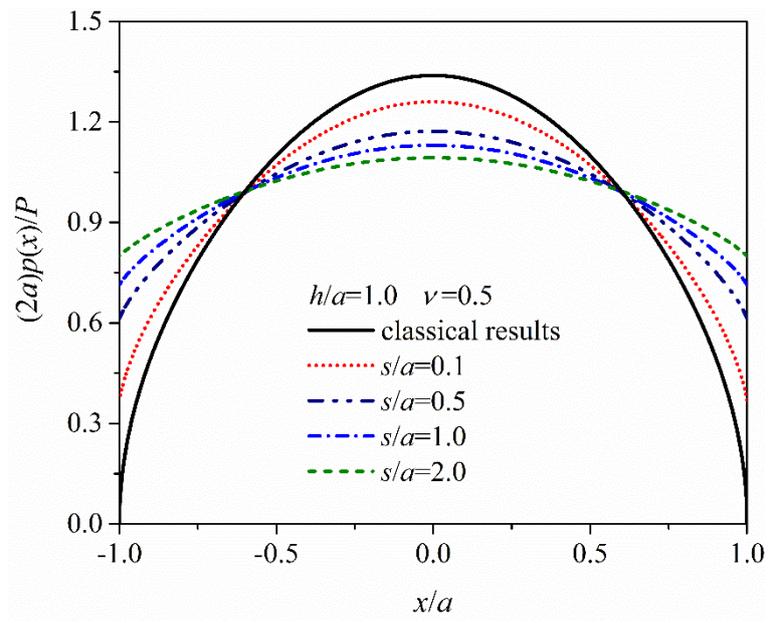

Figure 7

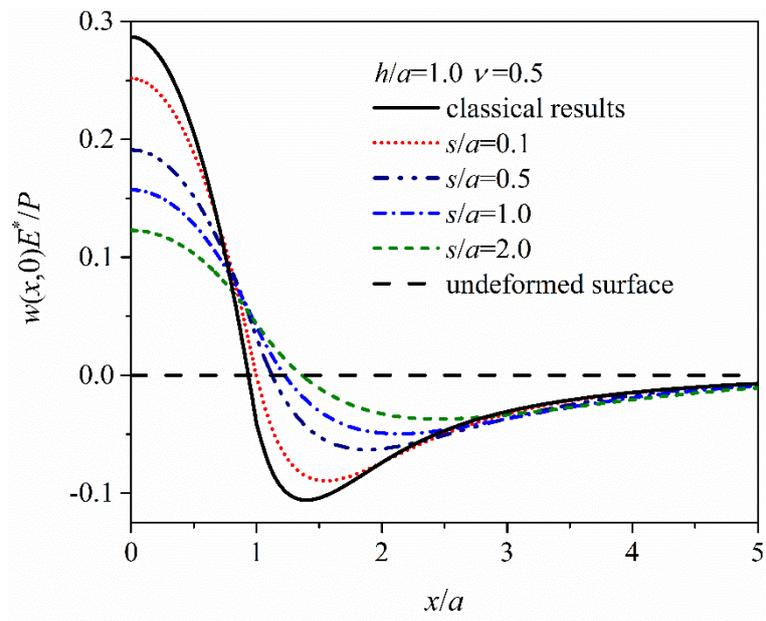

(a)

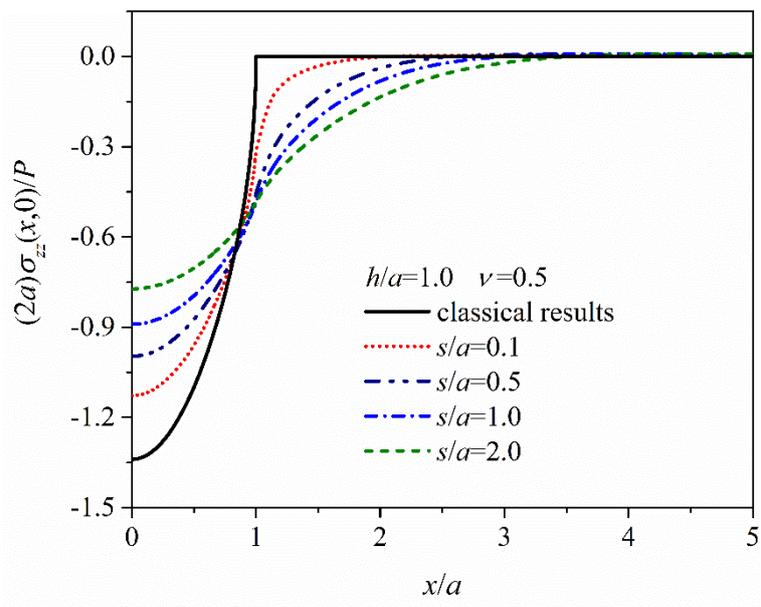

(b)

Figure 8

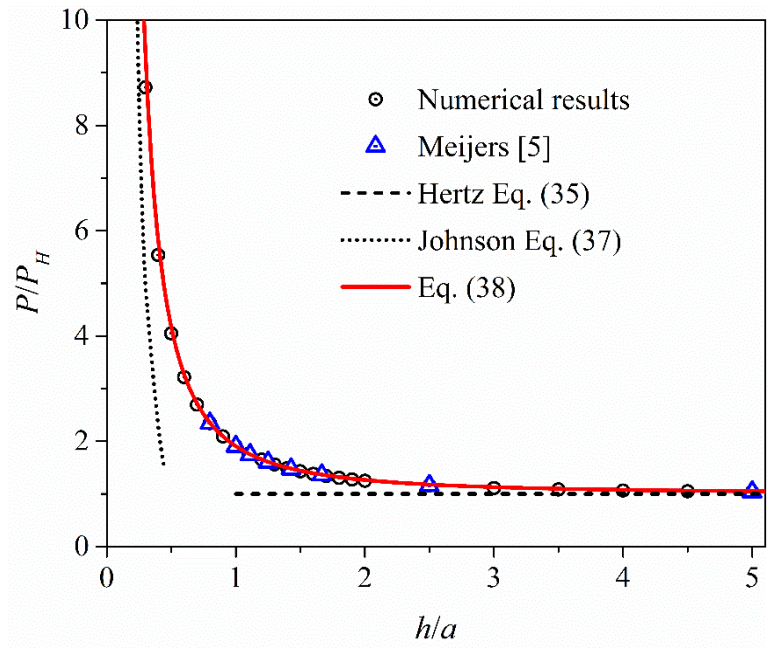

Figure 9

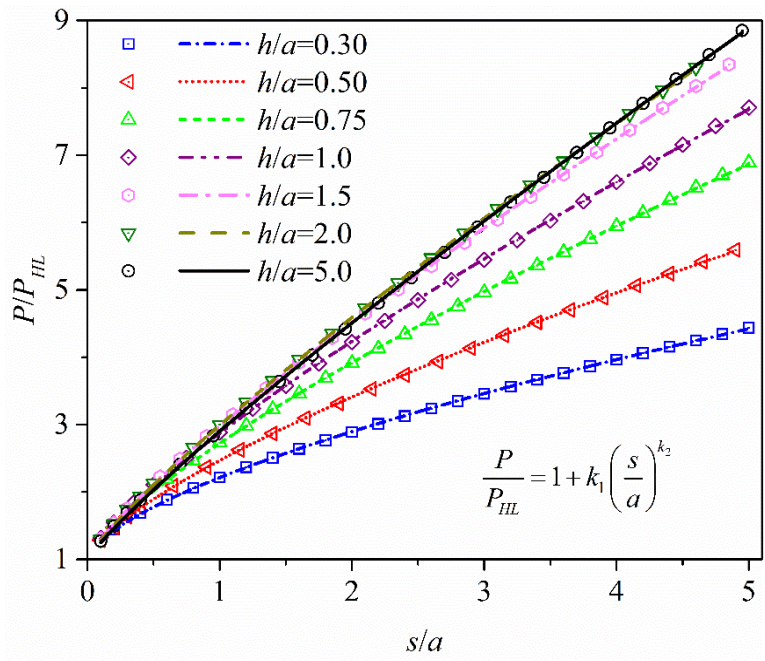

Figure 10

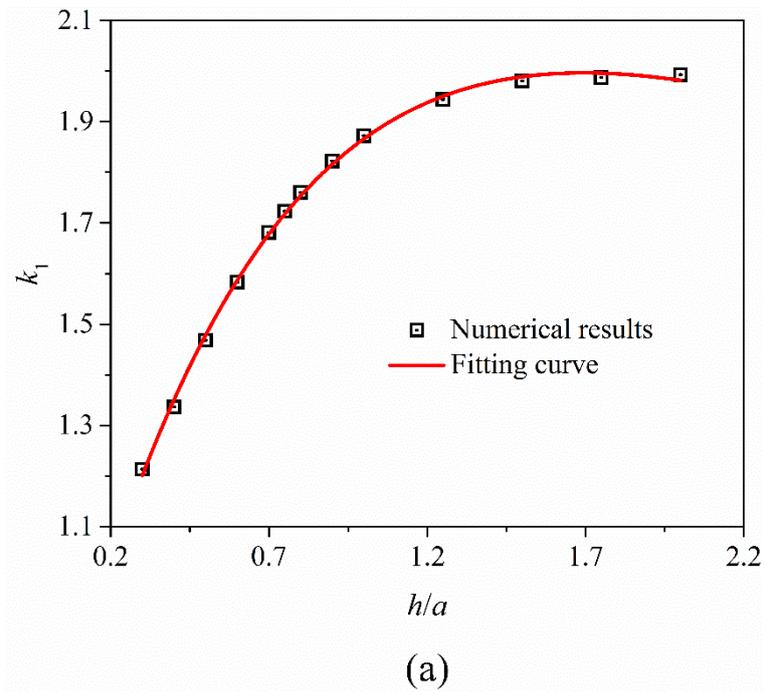

(a)

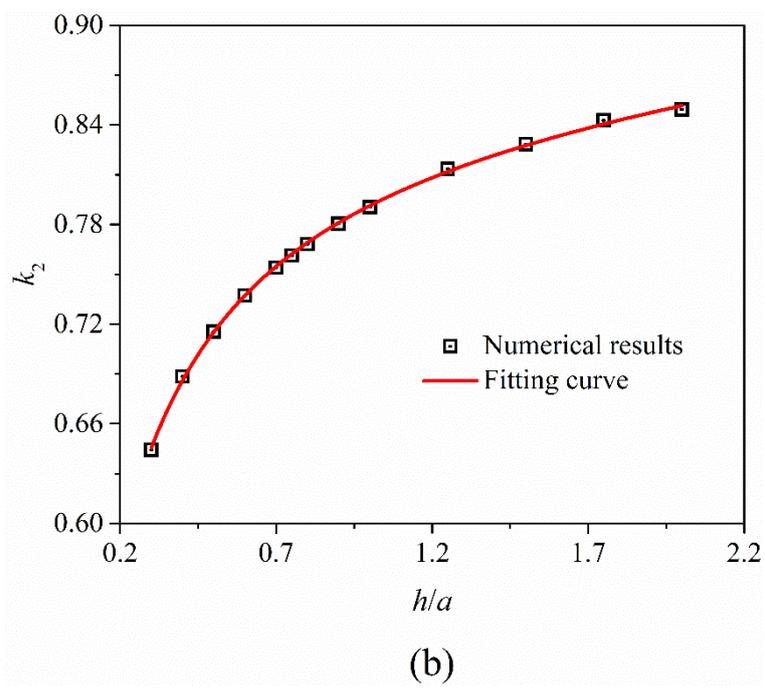

(b)

Figure 11